\documentstyle[12pt,worldsci]{article}
\begin{document}
\setlength{\baselineskip}{2.6ex}
\def\Re{{\cal R \mskip-4mu \lower.1ex \hbox{\it e}\,}}
\def\Im{{\cal I \mskip-5mu \lower.1ex \hbox{\it m}\,}}
\def\ie{{\it i.e.}}
\def\eg{{\it e.g.}}
\def\etc{{\it etc}}
\def\etal{{\it et al.}}
\def\ibid{{\it ibid}.}
\def\sub#1{_{\lower.25ex\hbox{$\scriptstyle#1$}}}
\def\sul#1{_{\kern-.1em#1}}
\def\sll#1{_{\kern-.2em#1}}
\def\sbl#1{_{\kern-.1em\lower.25ex\hbox{$\scriptstyle#1$}}}
\def\ssb#1{_{\lower.25ex\hbox{$\scriptscriptstyle#1$}}}
\def\sbb#1{_{\lower.4ex\hbox{$\scriptstyle#1$}}}
\def\spr#1{^{\,#1}}
\def\spl#1{^{\!#1}}
\def\tev{\,{\rm TeV}}
\def\gev{\,{\rm GeV}}
\def\mev{\,{\rm MeV}}
\def\to{\rightarrow}
\def\dmix{\ifmmode D^0-\bar D^0 \else $D^0-\bar D^0$\fi}
\def\dm{\ifmmode \Delta m_D \else $\Delta m_D$\fi}
\def\mh{\ifmmode m\sbl H \else $m\sbl H$\fi}
\def\mch{\ifmmode m_{H^\pm} \else $m_{H^\pm}$\fi}
\def\mt{\ifmmode m_t\else $m_t$\fi}
\def\mc{\ifmmode m_c\else $m_c$\fi}
\def\mz{\ifmmode M_Z\else $M_Z$\fi}
\def\mw{\ifmmode M_W\else $M_W$\fi}
\def\mws{\ifmmode M_W^2 \else $M_W^2$\fi}
\def\mhs{\ifmmode m_H^2 \else $m_H^2$\fi}
\def\mzs{\ifmmode M_Z^2 \else $M_Z^2$\fi}
\def\mts{\ifmmode m_t^2 \else $m_t^2$\fi}
\def\mcs{\ifmmode m_c^2 \else $m_c^2$\fi}
\def\mchs{\ifmmode m_{H^\pm}^2 \else $m_{H^\pm}^2$\fi}
\def\ztwo{\ifmmode Z_2\else $Z_2$\fi}
\def\zone{\ifmmode Z_1\else $Z_1$\fi}
\def\mtwo{\ifmmode M_2\else $M_2$\fi}
\def\mone{\ifmmode M_1\else $M_1$\fi}
\def\tb{\ifmmode \tan\beta \else $\tan\beta$\fi}
\def\xw{\ifmmode x\sub w\else $x\sub w$\fi}
\def\ch{\ifmmode H^\pm \else $H^\pm$\fi}
\def\lum{\ifmmode {\cal L}\else ${\cal L}$\fi}
\def\inpb{\ifmmode {\rm pb}^{-1}\else ${\rm pb}^{-1}$\fi}
\def\infb{\ifmmode {\rm fb}^{-1}\else ${\rm fb}^{-1}$\fi}
\def\epem{\ifmmode e^+e^-\else $e^+e^-$\fi}
\def\ppb{\ifmmode \bar pp\else $\bar pp$\fi}
\def\subw{_{\rm w}}
\def\half{\textstyle{{1\over 2}}}
\def\elli{\ell^{i}}
\def\ellj{\ell^{j}}
\def\ellk{\ell^{k}}
\newskip\zatskip \zatskip=0pt plus0pt minus0pt
\def\matth{\mathsurround=0pt}
\def\lsim{\mathrel{\mathpalette\atversim<}}
\def\gsim{\mathrel{\mathpalette\atversim>}}
\def\atversim#1#2{\lower0.7ex\vbox{\baselineskip\zatskip\lineskip\zatskip
  \lineskiplimit 0pt\ialign{$\matth#1\hfil##\hfil$\crcr#2\crcr\sim\crcr}}}
\def\undertext#1{$\underline{\smash{\vphantom{y}\hbox{#1}}}$}

\rightline{\vbox{\halign{&#\hfil\cr
&SLAC-PUB-6674\cr
&September 1994\cr
&T/E\cr}}}
\title{{\bf PROBING NEW PHYSICS IN RARE CHARM PROCESSES }
\footnote{Work Supported by the Department of Energy,
Contract DE-AC03-76SF00515}
\footnote{Presented at the {\it Meeting of the American Physical Society,
Division of Particles and Fields (DPF'94)}, Albuquerque, NM, August 2-6,
1994}}
\author{J.L.\ HEWETT\\
\vspace{0.3cm}
{\em Stanford Linear Accelerator Center, Stanford University, Stanford, CA
94309, USA}}
\maketitle

\begin{center}
\parbox{13.0cm}
{\begin{center} ABSTRACT \end{center}
{\small\hspace*{0.3cm}
The possibility of using the charm system to search for new physics is
addressed.  Phenomena such as $D^0-\bar D^0$ mixing and rare decays of charmed
mesons are first examined in the Standard Model to test our present
understanding and
to serve as benchmarks for signals from new sources.
The effects of new physics from various classes of non-standard dynamical
models on \dmix\ mixing are investigated.
}}
\end{center}

We examine the prospect of using one-loop processes in the charm sector
as a laboratory for probing new physics.  Similar processes in the $K$
system have played a strong and historical role\cite{kaon} in constraining
new physics, while corresponding investigations have just begun\cite{cleo}
in the b-quark sector.  In contrast, charm has played a lesser role in the
search for new physics. Due to the effectiveness of the GIM
mechanism, short distance Standard Model (SM) contributions to rare
charm processes are very small.  Most reactions are thus dominated by
long distance effects which are difficult to reliably calculate.
However, a recent estimation\cite{charmcrew} of such effects indicates
that there is a window for the clean observation of new physics in some
interactions.  In fact, it is precisely because the SM flavor changing neutral
current (FCNC) rates are so small, that charm provides an important opportunity
to discover new effects, and offers
a detailed test of the SM in the up-quark sector.  Due to space-time
limitations\cite{charmcrew}, this talk will concentrate on
\dmix\ mixing.  First, the SM predictions are reviewed, and then the
expectations in various extensions of the SM are discussed.  We conclude
with a brief summary of SM rates for rare $D$ meson decays.

Currently, the best limits\cite{pdg} on \dmix\ mixing are from fixed target
experiments, with $x_D\equiv\Delta m_D/\Gamma<0.083$ (where $\Delta
m_D=m_2-m_1$
is the mass difference), yielding $\dm<1.3\times 10^{-13}\gev$.
The bound on the ratio of wrong-sign to right-sign final
states is $r_D\equiv\Gamma(D^0\to\ell^-X)/\Gamma(D^0\to\ell^+X)<3.7\times
10^{-3}$, where
\begin{equation}
r_D\approx {1\over 2}\left[ \left( {\Delta m_D\over\Gamma}\right)^2 +
\left( {\Delta\Gamma\over 2\Gamma}\right)^2\right] \,,
\end{equation}
in the limit $\Delta m_D/\Gamma, \Delta\Gamma/\Gamma\ll 1$.
Several high volume charm experiments are planned for the future,
with $10^8$ charm mesons expected to be reconstructed.  Several rare
processes, including \dmix\ mixing, can then be probed another $1-2$ orders of
magnitude below present sensitivities.

The short distance SM contributions to \dm\ proceed through a $W$ box diagram
with internal $d,s,b$-quarks.  In this case the external momentum, which is
of order $m_c$, is communicated to the light quarks in the loop and
can not be neglected.  The effective Hamiltonian is
\begin{equation}
{\cal H}^{\Delta c=2}_{eff} = {G_F\alpha\over 8\sqrt 2\pi x_w}\left[
|V_{cs}V^*_{us}|^2 \left(I_1^s {\cal O}-m_c^2I_2^s {\cal O'}\right)+
|V_{cb}V^*_{ub}|^2\left( I_3^b {\cal O}-m_c^2I_4^b {\cal O'}\right) \right] \,,
\end{equation}
where the $I_j^{q}$ represent integrals\cite{datta} that are functions of
$m_{q}^2/M_W^2$ and $m_{q}^2/m_c^2$, and ${\cal O}=[\bar u\gamma_\mu
(1-\gamma_5)c]^2$ is the usual mixing operator while ${\cal O'}=[
\bar u(1+\gamma_5)c]^2$ arises in the case of non-vanishing external
momentum.  The numerical value of the short distance contribution is
$\dm\sim 5\times 10^{-18}$ GeV (taking $f_D=200\mev$).  The long distance
contributions have been computed via two different techniques: (i) the
intermediate particle dispersive approach yields\cite{charmcrew,gusto}
$\dm\sim 10^{-4}\Gamma\simeq 10^{-16}\gev$, and (ii) heavy quark
effective theory which results\cite{hqet} in $\dm\sim (1-2)\times 10^{-5}
\Gamma\simeq 10^{-17}\gev$.  Clearly, the SM predictions lie far below the
present experimental sensitivity!

One reason the SM expectations for \dmix\ mixing are so small is that there
are no heavy particles participating in the box diagram to enhance the rate.
Hence the first extension to the SM that we consider is the
addition\cite{four} of a heavy $Q=-1/3$ quark which may be present, \eg, as an
iso-doublet fourth generation $b'$-quark, or as a singlet quark in $E_6$ grand
unified theories.  The current bound\cite{pdg} on the mass of such an object
is $m_{b'}>85\gev$, assuming that it decays via charged current interactions.
We can now neglect the external momentum and \dm\ is given
by the usual expression\cite{il},
\begin{equation}
\dm={G_F^2M_W^2m_D\over 6\pi^2}f_D^2B_D|V_{cb'}V_{ub'}^*|^2F(m^2_{b'}/M_W^2)
\,.
\end{equation}
The value of \dm\ is displayed in this model in Fig.\ 1a as a function of the
overall CKM mixing factor for various values of the heavy quark mass.  We see
that \dm\ approaches the experimental bound for large values of the
mixing factor.  A naive estimate in the four generation SM yields\cite{pdg} the
restrictions $|V_{cb'}|<0.571$ and $|V_{ub'}|<0.078$.

Another simple extension of the SM is to enlarge the Higgs sector by an
additional doublet.  First, we examine two-Higgs-doublet models which
avoid tree-level FCNC by introducing a global symmetry.  Two such
models are Model I, where one doublet ($\phi_2$) generates masses for all
fermions and the second ($\phi_1$) decouples from the fermion sector, and
Model II, where $\phi_2$ gives mass to the up-type quarks, while the down-type
quarks and charged leptons receive their mass from $\phi_1$.  Each doublet
receives a vacuum expectation value $v_i$, subject to the constraint that
$v_1^2+v_2^2=v^2_{\rm SM}$.  The charged Higgs boson present in these models
will participate in the box diagram for \dm.  The $H^\pm$ interactions with
the quark sector are governed by the Lagrangian
\begin{equation}
{\cal L}={g\over 2\sqrt 2 M_W}H^\pm[V_{ij}m_{u_i}A_u\bar u_i(1-\gamma_5)d_j
+V_{ij}m_{d_j}A_d\bar u_i(1+\gamma_5)d_j]+h.c. \,,
\end{equation}
with $A_u=\cot\beta$ in both models and $A_d=-\cot\beta(\tan\beta)$ in Model
I(II), where $\tan\beta\equiv v_2/v_1$.
The expression for \dm\ in these models can be found in Ref. (9).
{}From the Lagrangian it is clear that Model I will only modify the SM
result for \dm\ for very small values of $\tan\beta$, and this region is
already
excluded\cite{cleo,bhp} from $b\to s\gamma$ and $B_d^0-\overline B_d^0$ mixing.
However, enhancements can occur in Model II for large values of
$\tan\beta$, as demonstrated in Fig.\ 1b.

Next we consider the case of extended Higgs sectors without natural flavor
conservation.  In these models the above requirement of a global symmetry
which restricts each fermion type to receive mass from only one doublet is
replaced\cite{fcnch} by approximate flavor symmetries which act on the
fermion sector.  The Yukawa couplings can then possess a structure which
reflects the observed fermion mass and mixing
hierarchy.  This allows the low-energy FCNC limits to be evaded as the
flavor changing couplings to the light fermions are small.  We employ the
Cheng-Sher ansatz\cite{fcnch}, where the flavor changing couplings of the
neutral Higgs are $\lambda_{h^0f_if_j}\approx (\sqrt 2G_F)^{1/2}
\sqrt{m_im_j}\Delta_{ij}$, with the $m_{i(j)}$ being the relevant fermion
masses and $\Delta_{ij}$ representing a combination of mixing angles.
$h^0$ can now contribute to \dm\ through tree-level exchange and the result is
displayed in Fig. 2a as a function of the mixing factor.
\dmix\ mixing can also be mediated by $h^0$ and t-quark virtual
exchange in a box diagram, however these contributions only compete with those
from the tree-level process for large values of $\Delta_{ij}$.  In Fig. 2b we
show the constraints placed on the parameters of this model from the present
experimental bound on \dm\ for both the tree-level and box diagram
contributions.

The last contribution to \dmix\ mixing that we will discuss here is that
of scalar leptoquark bosons.
Leptoquarks are color triplet particles which couple to a lepton-quark pair
and are naturally present in many theories beyond the SM which relate
leptons and quarks at a more fundamental level.  We parameterize their
{\it a priori} unknown couplings as $\lambda^2_{\ell q}/4\pi=F_{\ell q}\alpha$.
They participate in \dm\ via virtual exchange inside a box
diagram\cite{leptos}, together with a charged lepton or neutrino.
Assuming that there is no leptoquark-GIM mechanism, and taking both exchanged
leptons to be the same type, we obtain the restriction
\begin{equation}
{F_{\ell c}F_{\ell u}\over m^2_{LQ}} <
{196\pi^2 \dm\over (4\pi\alpha f_D)^2m_D} \,.
\end{equation}
The resulting bounds in the leptoquark coupling-mass plane are presented
in Fig. 3.

We close our discussion by displaying the expected branching
fractions for various rare charm decay modes in the SM in Table 1.
We present both the short distance predictions
(neglecting QCD corrections, which may be important in some decay modes),
an upper bound on the long distance estimates, as well as the
current experimental limits\cite{pdg,raredk}.  For more details we refer the
reader to Ref. (3).  We urge our experimental colleagues to
continue the search for rare charm processes!

\vspace{1.0cm}
%
\def\MPL #1 #2 #3 {Mod.~Phys.~Lett.~{\bf#1},\ #2 (#3)}
\def\NPB #1 #2 #3 {Nucl.~Phys.~{\bf#1},\ #2 (#3)}
\def\PLB #1 #2 #3 {Phys.~Lett.~{\bf#1},\ #2 (#3)}
\def\PR #1 #2 #3 {Phys.~Rep.~{\bf#1},\ #2 (#3)}
\def\PRD #1 #2 #3 {Phys.~Rev.~{\bf#1},\ #2 (#3)}
\def\PRL #1 #2 #3 {Phys.~Rev.~Lett.~{\bf#1},\ #2 (#3)}
\def\RMP #1 #2 #3 {Rev.~Mod.~Phys.~{\bf#1},\ #2 (#3)}
\def\ZP #1 #2 #3 {Z.~Phys.~{\bf#1},\ #2 (#3)}
\def\IJMP #1 #2 #3 {Int.~J.~Mod.~Phys.~{\bf#1},\ #2 (#3)}
\bibliographystyle{unsrt}

\begin{table}
\centering
\begin{tabular}{|l|c|c|c|} \hline\hline
Decay Mode & Experimental Limit & $B_{S.D.}$ & $B_{L.D.}$ \\ \hline\hline
$D^0\to\mu^+\mu^-$ & $<1.1\times 10^{-5}$ & $(1-20)\times 10^{-19}$ &
$<3\times 10^{-15}$ \\
$D^0\to\mu^\pm e^\mp$ & $<1.0\times 10^{-4}$ & $0$ & $0$ \\ \hline
$D^0\to\gamma\gamma$ & --- & $10^{-16}$ & $<3\times 10^{-9}$ \\ \hline
$D\to X_u+\gamma$ & & $1.4\times 10^{-17}$ & \\
$D^0\to\rho^0\gamma$ & $<1.4\times 10^{-4}$ & & $<2\times 10^{-5}$ \\
$D^0\to\phi^0\gamma$ & $<2.0\times 10^{-4}$ & & $<10^{-4}$ \\
$D^+\to\rho^+\gamma$ & --- & & $<2\times 10^{-4}$ \\ \hline
$D\to X_u+\ell^+\ell^-$ & & $4\times 10^{-9}$ & \\
$D^0\to\pi^0\mu\mu$ & $<1.7\times 10^{-4}$ & & \\
$D^0\to\bar K^0 ee/\mu\mu$ & $<17.0/2.5\times 10^{-4}$ & &
$<2\times 10^{-15}$ \\
$D^+\to\pi^+ee/\mu\mu$ & $<250/4.6\times 10^{-5}$ & few$\times 10^{-10}$ &
$<10^{-8}$ \\
$D^+\to K^+ee/\mu\mu$ & $<480/8.5\times 10^{-5}$ & & $<10^{-15}$ \\
\hline
$D^0\to X_u+\nu\bar\nu$ & & $2.0\times 10^{-15}$ & \\
$D^0\to\pi^0\nu\bar\nu$ & --- & $4.9\times 10^{-16}$ & $<6\times 10^{-16}$ \\
$D^0\to\bar K^0\nu\bar\nu$ & --- & & $<10^{-12}$ \\
$D^+\to X_u+\nu\bar\nu$ & --- & $4.5\times 10^{-15}$ & \\
$D^+\to\pi^+\nu\bar\nu$ & --- & $3.9\times 10^{-16}$ & $<8\times 10^{-16}$ \\
$D^+\to K^+\nu\bar\nu$ & --- & & $<10^{-14}$ \\ \hline\hline
\end{tabular}
\caption{Standard Model predictions for the branching fractions due to short
and long distance contributions for various rare $D$ meson decays. Also
shown are the current experimental limits.}
\end{table}

\newpage

\noindent
Fig.~1: \dm\ in (a) the four generation SM as a function of the CKM mixing
factor with the solid, dashed, dotted, dash-dotted curve corresponding to
 $m_{b'}=100, 200, 300, 400$ GeV, respectively. (b) in two-Higgs-doublet model
II as a function of $\tan\beta$ with, from top to bottom, the solid, dashed,
dotted, dash-dotted, solid curve representing $m_{H^\pm}=50, 100, 250, 500,
1000\gev$.  The solid horizontal line corresponds to the present experimental
limit.

\bigskip

\noindent
Fig.~2: (a) \dm\ in the flavor changing Higgs model described in the text
as a function of the mixing factor with $m_{h^0}=50, 100, 250, 500, 1000\gev$
corresponding to the solid, dashed, dotted, dash-dotted, solid curve from
top to bottom.  (b)  Constraints in the mass-mixing factor plane from \dm\
from the tree-level process (solid curve) and the box diagram (dashed).

\bigskip

\noindent
Fig.\ 3:  Constraints in the leptoquark coupling-mass plane from \dm.

\end{document}